%% file: dense_pqcd_PRL.tex
\documentclass[aps,prl,twocolumn,superscriptaddress,showpacs,floatfix,preprintnumbers]{revtex4-1}

\usepackage{hyperref}
\usepackage{color}
\usepackage{graphicx}	% Include figure field
\graphicspath{{./figures/}}
\usepackage[utf8]{inputenc}
\usepackage{amsmath} \usepackage{amssymb}

\allowdisplaybreaks

\renewcommand{\vec}{\mathbf}
\newcommand{\order}[1]{\mathcal{O}( #1 )}
\newcommand{\dOne}[1]{\operatorname{d} \! #1}
\newcommand{\dDim}[2]{\!\frac{\operatorname{d}^{#1} \! #2}{(2\pi)^{#1}}}
\newcommand{\imag}{\mathrm{i}}

\renewcommand{\vec}{\boldsymbol}

\newcommand{\pilp}{\Pi_L\left(\Phi_P\right)}
\newcommand{\pilq}{\Pi_L\left(\Phi_Q\right)}
\newcommand{\pitp}{\Pi_T\left(\Phi_P\right)}
\newcommand{\pitq}{\Pi_T\left(\Phi_Q\right)}

\begin{document}
\title{NNNLO pressure of cold quark matter: leading logarithm}

\preprint{CERN-TH-2018-230, HIP-2018-13/TH}
\author{Tyler Gorda}
\affiliation{Department of Physics, University of Virginia, Charlottesville, VA, USA}
\affiliation{Helsinki Institute of Physics and Department of Physics, University of Helsinki, Finland}
\author{Aleksi Kurkela}
\affiliation{Theoretical Physics Department, CERN, Geneva, Switzerland} 
\affiliation{Faculty of Science and Technology, University of Stavanger, Stavanger, Norway}
\author{Paul Romatschke}
\affiliation{Department of Physics, University of Colorado Boulder, Boulder, CO, USA}
\affiliation{Center for Theory of Quantum Matter, University of Colorado, Boulder, CO, USA}
\author{Saga S\"appi}
\affiliation{Helsinki Institute of Physics and Department of Physics, University of Helsinki, Finland}
\author{Aleksi Vuorinen}
\affiliation{Helsinki Institute of Physics and Department of Physics, University of Helsinki, Finland}

\begin{abstract}
At high baryon chemical potential $\mu_B$, the equation of state of QCD allows a weak-coupling expansion in the QCD coupling $\alpha_s$. The result is currently known up to and including the full next-to-next-to-leading order (NNLO) $\alpha_s^2$. Starting at this order, the computations are complicated by the modification of particle propagation in a dense medium, which necessitates non-perturbative treatment of the scale $\alpha_s^{1/2}\mu_B$. In this work, we apply a Hard-Thermal-Loop scheme for capturing the contributions of this scale to the weak-coupling expansion, and use it to determine the leading-logarithm contribution to N$^3$LO: $\alpha_s^3 \ln^2 \alpha_s$. This result is the first improvement to the equation of state of massless cold quark matter in 40 years. The new term is negligibly small, and thus significantly increases our confidence in the applicability of the weak-coupling expansion.
\end{abstract}

\maketitle

\emph{Introduction}.---Quantum Chromodynamics (QCD) is the accepted theory of the strong interaction, and describes a wide range of physical phenomena from the masses and properties of hadrons to the observable characteristics of neutron stars. In the limit of high density, the theory is, however, notoriously difficult to solve, as lattice simulations are plagued by the infamous sign problem (for some approaches to overcome it, see, e.g., refs.~\cite{Aarts:2013lcm,Fodor:2015doa,Cristoforetti:2012su,Fujii:2013sra,deForcrand:2014tha,Ichihara:2015kba,Alexandru:2015sua}). In the limit of very high densities, the asymptotic freedom of QCD \cite{PhysRevLett.30.1343} suggests that a weak-coupling approach to the thermodynamics of the deconfined phase, i.e.,~quark matter, might be feasible, but in practice the application of perturbation theory is very challenging. In fact, no new perturbative orders have been determined for the equation of state (EoS) since 1977, when Freedman and McLerran derived the full next-to-next-to-leading order (NNLO) result for the pressure as a function of quark chemical potentials in the limit of massless quarks \cite{Freedman:1976dm,Freedman:1976ub}. Since then, this result has been generalized to the $\overline{\text{MS}}$ scheme \cite{Vuorinen:2003fs}, to include finite temperature effects \cite{Ipp:2006ij,Kurkela:2016was}, and to nonzero quark masses \cite{Fraga:2004gz,Kurkela:2009gj,Fraga:2013qra}, but no realistic attempts to reach N$^3$LO have been made so far.

In a strongly coupled medium at large baryonic density, interactions with the medium constituents lead to the screening of color charges---a phenomenon that is a nonabelian generalization of Debye screening. This generates a new in-medium mass scale $m_\infty \sim \alpha_{s}^{1/2} \mu_{B} \ll \mu_B$, a scale which we shall refer to as ``soft''. Here $\alpha_{s}$ is the strong coupling constant and $\mu_{B}$ the baryon number chemical potential \footnote{Note that although the pressure is in principle a function of several independent quark chemical potentials, in this work we will consistently parameterize it in terms of the single baryon chemical potential $\mu_B$. The reason for this stems from the fact that with three massless quarks, the physically relevant limits of local charge neutrality and $\beta$-equilibrium are satisfied when $\mu_u=\mu_d=\mu_s=\mu_B/3$. Note, however, that it is trivial to generalize our result to the case of unequal quark chemical potentials, as the new term in the pressure only depends on them via the $m_\infty$ parameter.}. This new scale manifests as infrared (IR) divergences in naive loop expansions, and a proper handling of the soft sector to a given order in $\alpha_s$ requires a resummation of diagrams with an arbitrary number of loops. In this sense, the soft scale requires non-perturbative treatment.  These non-perturbative effects predominantly arise through interactions of the soft modes with the typical modes in the medium, which have momenta proportional to $\mu_B$, a scale which we shall refer to as ``hard''. Due to the small number of soft modes, the interactions among the soft modes amount to a subdominant perturbative correction. Diagrammatically, this is reflected in the restricted set of topologies that require special treatment, namely only soft gluonic propagators and vertex functions need to be resummed. 

While the naive loop expansion of the EoS leads to a series of terms analytic in $\alpha_s$, this need not be the case for the resummed soft sector: In particular, loop integrals that are sensitive to both the hard and the soft scales can also receive contributions from the \emph{semisoft} region between the two. This leads to logarithms of the ratio of the scales ${\int_{\alpha_s^{1/2} \mu_B}^{\mu_B} \dOne^4 P/(P^2)^2 \sim \ln (\alpha_{s}^{1/2} \mu_B / \mu_B})$, and gives rise to non-analytic terms in the weak coupling series (here and in what follows, $P$ denotes the magnitude of a Euclidean four-vector). The first order at which these non-analytic terms appear is NNLO, where they lead to a term proportional to $\alpha_{s}^{2} \ln \alpha_{s}$, derived in refs.~\cite{Freedman:1976dm,Freedman:1976ub,Kurkela:2016was}. 

As shall be made clear in this letter, the pressure $p$ of cold and dense three-color, three-flavor ($N_{c} = N_{f} = 3$) QCD matter with massless quarks can be written in the form (see, e.g.,\ \cite{Kurkela:2016was})
\begin{eqnarray}
    p\!\!&\simeq&\!\! \frac{3 (\mu_B/3)^4}{4\pi^2}\bigg[    1 - 0.636620 \,\alpha_{s} - 0.303964 \,\alpha_{s}^{2} \ln \,\alpha_{s}\nonumber\\
    &&- \left(0.874355 + 0.911891 \ln \frac{\bar{\Lambda}}{\mu_B/3}\right) \alpha_{s}^{2}\bigg] \label{eq:p_expansion} \\
    &&+ c_{3,2} \,\alpha_{s}^{3} \ln^2 \alpha_{s} + c_{3,1}(\bar{\Lambda}) \alpha_{s}^{3} \ln \alpha_{s} + c_{3,0}(\bar{\Lambda}) \alpha_{s}^{3} +\order{\alpha_{s}^{4}},\nonumber 
\end{eqnarray}
where $\bar\Lambda$ is the renormalization scale, and where the $c_{3,i}$ are the as-yet-uncalculated N$^{3}$LO terms. In this work, we apply the methodology of separating the soft contributions to the pressure presented in ref.~\cite{Kurkela:2016was}, which allows us to cleanly separate the logarithmic terms in the expansion. This methodology is used to determine the first fundamentally new perturbative order in the EoS since the Freedman--McLerran calculation: We shall calculate the coefficient $c_{3,2}$ in the equation above, which gives the dominant N$^3$LO contribution in the $\alpha_s \rightarrow 0$ limit. 

Besides a purely theoretical interest in the problem, there is strong motivation stemming from a hope that new perturbative orders will decrease the systematic uncertainty in the EoS in a range of densities where it might be relevant for the physics of neutron stars. Indeed, it has recently been demonstrated that the EoS of neutron-star matter can be significantly constrained by combining first-principles information from both low and high densities with astrophysical observations \cite{Kurkela:2014vha,Gorda:2016uag,Annala:2017llu,Most:2018hfd}. In light of the present emergence of the discipline of gravitational wave astronomy, there is a real prospect that an active interplay between QCD calculations, numerical relativity, and observations will provide a way to deepen our understanding of how nature works in a previously inaccessible domain \cite{TheLIGOScientific:2017qsa}.

\emph{Warm-up computation and setup}.---Let us start by briefly considering how the leading non-analytic term of $\order{\alpha_{s}^{2}\ln \alpha_{s}}$ enters the weak-coupling expansion of the QCD pressure at $T = 0$. At leading order $\alpha_s^0$, the gluonic contribution to the pressure is given by the simple vacuum diagram 
\begin{equation}
    \vcenter{\hbox{\includegraphics{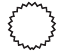}}}\!  = -\frac{d_A}{2} \! \int \! \dDim{4}{P} \bigg[(2+1) \ln \left( P^2 \right) \bigg],
\label{eq:bare}
\end{equation}
where the (2+1) corresponds to 2 transverse polarization modes and one longitudinal (eventually removed by a ghost diagram, not shown).
While this gluonic contribution is divergent, it is clearly independent of $\mu_{B}$ and in fact vanishes upon vacuum $\mu_B = 0$ subtraction. 
(The corresponding fermionic 1-loop diagram gives the Fermi--Dirac pressure of free quarks.) Corrections at higher orders in $\alpha_s$ arise from decorating the above diagram with an increasing number of propagators. If the momentum flowing in all the lines is of order $\mu_B$, this gives rise to the naive loop expansion. However, when the integration momentum in eq.~\eqref{eq:bare} becomes soft, $P \sim \alpha_s^{1/2} \mu_B$, adding an arbitrary number of (one-loop) self-energy insertions to the gluon line does not change the magnitude of the diagram. Therefore the naive loop expansion gets the answer wrong by an amount $\sim \int_0^{\alpha_s^{1/2}\mu_B} \dOne^4 P \sim \alpha_s^2 \mu_B^4$, which can be corrected by removing the naive expression in the relevant kinematic regime  (i.e.,~by introducing a counterterm) and by adding the resummed two-particle-reducible (2PR) ``ring diagrams'' of refs.~\cite{Freedman:1976dm,Freedman:1976ub,Kajantie:2001hv} to the expression.

Unlike gluons, fermions are not sensitive to the soft scale. Only excitations above the hard Fermi momentum $p_{F} = \mu_{B}/3$ exist, as the softer fermions are Pauli blocked. Therefore at $T=0$ the fermionic 1-loop diagram, and more generally fermionic lines, do not require a similar treatment and instead give rise to a naive expansion in powers of $\alpha_s$. In addition, ghosts do not require resummations \cite{Braaten:1989mz}. 

%%%%%%%%%%%%%%%%%%%%%%%%%%%%%%%%%%%%%%%%%%%%%%%%%%%%%
\begin{figure}
\vspace{-0.2cm}
\includegraphics[width=0.39\textwidth]{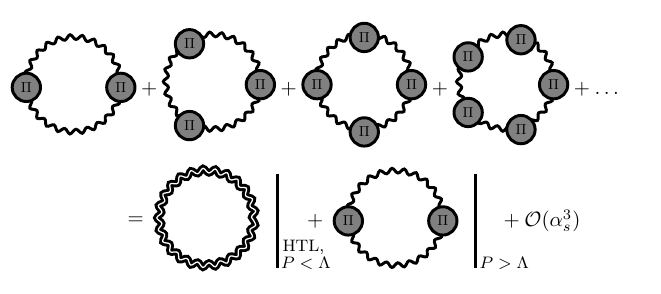}
\vspace{-0.5cm}
\caption{Graphical representation of eq.~\eqref{eq:ir_split} after kinematic simplifications are employed. See main text for explanation.}
\label{fig:ring_sum}
\end{figure}
%%%%%%%%%%%%%%%%%%%%%%%%%%%%%%%%%%%%%%%%%%%%%%%%%%%%%%

Consider now the resummed 1-loop ring sum depicted on the first line of Fig.~\ref{fig:ring_sum}. Since only the modes much softer than $\mu_{B}$ require resummation, we may split the integral over the loop momentum $P$ connecting the self-energy insertions into two regions by introducing a cutoff $\alpha_{s}^{1/2}\mu_B \ll \Lambda \ll \mu_{B}$, and revert to a naive loop expansion in the region $P > \Lambda$,
\begin{align}
p_{\text{IR}, 1}^{\text{res}} 
&= p_{\text{IR}, 1}^{\text{res}}(\{0,\Lambda\}) + p_{\text{IR}, 1}^{\text{loop}}(\{\Lambda, \infty \}), 
\label{eq:ir_split}
\end{align}
where the notation $\{.,.\}$ indicates the momentum cutoffs used. The momentum flowing in the self-energy insertions of $p_{\text{IR},1}^{\text{res}}$ may either be soft or hard.  If it is hard, then kinematic approximations may be employed and the self-energies can be expanded for small external momenta. To the leading order in the external momenta this gives rise to the well-known Hard Thermal Loop (HTL) power counting \cite{Braaten:1989mz}, and allows for a convenient computation of the resummed diagrams within the framework of the HTL effective theory \cite{Andersen:1999fw,Andersen:1999sf,Andersen:2002ey,Andersen:2003zk,Haque:2014rua}. On the other hand, if the momentum flowing in the self energy is soft, then this line (if it is gluonic) also needs to be resummed. However, because of the small volume of phase space, this contribution is subleading in $\alpha_s$. As we will see later, it is exactly these latter terms that give rise to the contributions we are after at N$^{3}$LO. 

The logarithmic contributions to the pressure arise from scaleless integrals in the \emph{semisoft} region $P \sim \Lambda$ between the soft and the hard scales, 
$
{\int_{\alpha^{1/2}\mu_B}^{\mu_B} \dOne^4P/P^4 \sim \ln \alpha_s^{1/2}},
$
and the coefficient of the leading NNLO logarithm can be extracted equally from the ultraviolet (UV) limit of $p^{\text{res}}_{\text{IR}, 1}$ or from the IR limit of $p^{\text{loop}}_{\text{IR}, 1}$. The semisoft contribution to the pressure is in fact particularly simple, as the propagator can be treated as if it were both soft and hard: Because $P\ll \mu_B$, instead of all topologies only the restricted HTL set of diagrams contribute, but because $P \gg \alpha_s^{1/2} \mu_B$ the diagram can be expanded in the number of self-energy insertions. 

To explicitly verify the above claims, we begin from the UV-regulated LO HTL pressure with the bare counter term (\ref{eq:bare}) subtracted \cite{Andersen:1999sf},
\begin{eqnarray}
\label{eq:1l_gluon_nonhard}
p_{\text{IR},1}^{\mathrm{HTL}} \!\!&=&\!
 \left[ \vcenter{\hbox{\includegraphics[width=1.0cm]{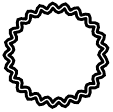}}}\, - \!\vcenter{\hbox{\includegraphics[width=1.20cm]{1-loop_gluon.pdf}}}\!\!\! \right] \\[0.1ex]
&=& -\frac{d_A}{2} \! \int^{\Lambda} \! \dDim{4}{P} \Bigg[  2 \ln \left( 1+ \frac{\Pi_T}{P^2} \right) + \ln \left( 1+ \frac{\Pi_L}{P^2} \right) \Bigg]\,,\nonumber  
  \end{eqnarray}
where the double line corresponds to the HTL-resummed propagator. Here, the longitudinal and transverse self energies read, in $d=3$ spatial dimensions,
\begin{align}
    \Pi_{L}\left(P\right) =&\,  2m_{\infty}^{2} \frac{P^{2}}{|\vec{p}|^{2}} \left[1-\frac{\imag P^{0}}{2\left|\vec{p}\right|}\ln\frac{\imag P^{0}+\left|\vec{p}\right|}{\imag P^{0}-\left|\vec{p}\right|}\right] \label{eq:pi_L},  \\
    \Pi_{T}\left(P\right) =&\, m_{\infty}^{2} - \frac{\Pi_{L}\left(P\right)}{2}, \label{eq:pi_T}
\end{align}
where $m_{\infty}^{2} = \alpha_{s} \mu_{B}^{2} N_{f} / (9 \pi)$ is the asymptotic HTL mass \cite{Andersen:1999sf}. Note that the breaking of Lorentz symmetry originates from the rest frame singled out by the presence of the medium. 
Concentrating now on the semisoft region, we expand the logarithms in eq.~\eqref{eq:1l_gluon_nonhard} in powers of the self-energy. Introducing two semisoft momentum-space cutoffs $\alpha_{s}^{1/2} \mu_B \ll \Lambda_1 \ll \Lambda_2 \ll \mu_B $, we are left with the integral
\vspace{-0.2cm}
\begin{align}
    &p_{\text{IR},1}^{\mathrm{semisoft}} = -d_A \! \int_{\Lambda_1}^{\Lambda_2} \! \frac{\mathrm{d}^{4} P}{(2 \pi)^{4}} \left[ \frac{\Pi_{T} + \frac{\Pi_{L}}{2}}{P^2} - \frac{\Pi_{T}^{2} + \frac{\Pi_{L}^{2}}{2}}{2P^4} + \ldots \right] \nonumber \\
    &= -\frac{d_A}{(4 \pi)^2} \Bigg[ m_\infty^2 (\Lambda_2^2-\Lambda_1^2) - m_\infty^4 \ln \frac{\Lambda_2}{\Lambda_1}  
    + \order{\alpha_{s}^{3}} \Bigg]\,.
\label{eq:1l_gluon_semisoft}
\end{align}
The terms with a power-like dependence on the cutoffs $\Lambda_1$ and $\Lambda_2$ must cancel against corresponding terms
in $p^{\text{res}}_{\text{IR}, 1}(\{0,\Lambda_1\})$ and $p^{\text{loop}}_{\text{IR},1}(\{\Lambda_2,\infty\})$, respectively. Similarly, in the full expression the cutoff dependence in the logarithm is cancelled and $\Lambda_1$ and $\Lambda_2$ are replaced with quantities of magnitudes $\order{\alpha_s^{1/2}\mu_B}$ and $\order{\mu_B}$, as these are the only scales appearing in the soft and hard calculations. This gives the logarithm of $\alpha_s$ in the NNLO result.

There are two things to note about the calculation presented above. First, while the logarithmic term could be extracted from the semisoft region alone, obtaining the constant under the logarithm requires a precise calculation in both the hard and soft kinematic regions, which is a considerably more challenging task. Second, it turns out that the term non-analytic in $\alpha_s$ is the same as what one would obtain by setting the momentum $P$ on shell, with $\Pi_T = \Pi_T({\rm i} P_0 = | {\vec p}|, {\vec p} )= m_\infty^2$ and $\Pi_{L}=\Pi_L({\rm i} P_0 = | {\vec p}|, {\vec p} )= 0$, that is, by considering two massive transverse polarizations of gluons in the semisoft region. This is natural because this is the particle content of the HTL theory in its UV limit \cite{Weldon:1982aq}.

\emph{Applying the setup to \texorpdfstring{$\alpha_{s}^{3} \ln^2 \alpha_{s}$}{alpha\^{}3 ln\^{}2 alpha}}.---We have seen above how the single $\ln \alpha_s$ term in the NNLO pressure arises from a single semisoft integral. Similarly, if a diagram has multiple semisoft integrals, it has the potential to give rise to a higher power $\ln^{n} \alpha_s$. In particular, going to N$^3$LO we may allow two gluon lines in a given Feynman diagram to be soft, which opens up the possibility of obtaining a $\ln^2 \alpha_s$ term.

At N$^{3}$LO, there are three types of contributions to consider: Higher-order interactions between hard modes and other hard modes, higher-order interactions between soft modes and hard modes, and the first interactions between soft modes and other soft modes. Diagrammatically, the first arise from unresummed 4-loop diagrams, the second arise from single multi-loop 2PR self-energy insertions into the resummed diagrams in Fig.~\ref{fig:ring_sum}, and the last correspond to soft limits of resummed multi-loop vacuum diagrams.

The determination of the full N$^{3}$LO pressure is a daunting task. However, a full accounting of the different contributions listed above is not necessary in order to extract the leading-logarithm term at N$^{3}$LO, for the following reason. The insertion of a new soft loop to a soft line contributes a factor $\alpha_{s} \int \dOne^{4} P / P^{2} / m^{2}_{\infty} = \order{\alpha_{s}}$, where the factor $\alpha_{s}$ originates from the new vertex, $\int \dOne^{4} P / P^{2}$ from the loop integral and the inserted line, and $1/m_{\infty}^{2}$ from splitting the original soft propagator into two. This implies that the interactions of more than two soft momenta go beyond N$^{3}$LO. Therefore, the proper generalization of eq.~\eqref{eq:ir_split} to the N$^{3}$LO case will keep track of exactly two (gluonic) momenta. Introducing two semisoft scales $\alpha^{1/2}\mu_B \ll \Lambda^{i}   \ll \mu_{B}$, with $i=P,Q$, we thus have 
\begin{align}
p_{\text{IR, 2}}^{\text{res}} =&\,\,p_{\text{IR, 2}}^{\text{loop,P; loop,Q}}(\{\Lambda^{P},\infty\},\{\Lambda^{Q},\infty\})\nonumber \\
&+ p_{\text{IR, 2}}^{\text{res,P; loop,Q}}(\{0,\Lambda^{P}\},\{\Lambda^{Q},\infty\}) \nonumber \\
&+ p_{\text{IR, 2}}^{\text{loop,P; res,Q}}(\{\Lambda^{P},\infty\},\{0,\Lambda^{Q}\}) \nonumber \\ 
&+ p_{\text{IR, 2}}^{\text{res,P; res,Q}}(\{0,\Lambda^{P}\},\{0,\Lambda^{Q}\}).
\label{eq:breakdown}
\end{align}
Again the logarithms may be extracted from the $\Lambda^i$ dependence of the individual terms. The last term corresponds to a doubly-soft contribution, reproduced faithfully by the HTL resummation. In the second and third terms, one of the loop momenta is hard, so that the kinematic HTL approximation is insufficient, and additional diagrams that go beyond HTL must be considered. Finally, the first term corresponds to naive 4-loop (hard) diagrams, where no resummations are needed; these graphs are tabulated in ref.~\cite{Kajantie:2001hv}.

As in the NNLO case, the leading logarithm may be extracted from multiple places in the above expression. We choose to extract the double logarithm from the last term, as it corresponds to a previously-known two-loop HTL computation. Specifically, eq.~(34) of ref.~\cite{Andersen:2002ey} gives the integral expression for the gauge-invariant sum of the HTL-resummed diagrams 
\vspace{0cm}
\begin{equation}
 p^{\rm HTL}_{\rm IR,2} =
  \!\vcenter{\hbox{\includegraphics[width=5.35cm]{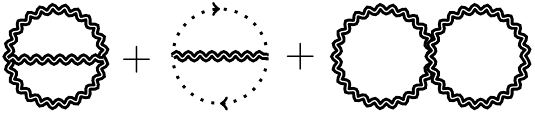}}} \, .
\end{equation}
In analogy to the previous section, we may expand this expression in the (now doubly-) semisoft limit to extract the leading $\ln^2 \alpha_s$ term: This amounts to an expansion in powers of $m_\infty^2$ to isolate the $m_\infty^4$ term, as it contains dimensionless integrals that can yield the double logarithm. \footnote{Note that this amounts to one further term in the expansion in $m_{D}^{2}$ than what was done in ref.~\cite{Andersen:2002ey}.}

Furthermore, to obtain the double logarithm, we need the two integration momenta to be well separated to produce scale-free integrals. Since $m_{\infty}^{4}$ already has the correct mass dimension for the pressure, we may rewrite the expanded HTL expression in the form
\begin{equation}
  \alpha_s  m_{\infty}^{4} \int \frac{\dOne^{4} P}{P^{4}} \frac{\dOne^{4} Q}{Q^{4}} f\! \left( \frac{P}{Q} , \Omega_{i} \right),
\label{eq:dimesionless}
\end{equation}
where the function $f$ is dimensionless, and inside the $f$ function $P$ and $Q$ represent the magnitudes of the Euclidean four-momenta and $\Omega_{i}$ represents the remaining angles. We have chosen to make the dimensionful denominator $P^{4} Q^{4}$, since we wish to extract precisely the integrals 
\begin{equation}
\int_{\Lambda_1^P}^{\Lambda_2^P} \int_{\Lambda_1^Q}^{\Lambda_2^Q} \frac{\mathrm{d}^4 P}{P^{4}}\frac{ \mathrm{d}^4 Q}{Q^{4}} \sim \ln^2 \alpha_{s}^{1/2} + \order{\ln \alpha_{s}, 1},
\label{eq:double_log_ints}
\end{equation} 
where the new semisoft cutoffs $\Lambda_{1,2}^{P},\Lambda_{1,2}^{Q}$ inside the $f$ function are defined as before. Analogously to the NNLO case, the double logarithm in the full expression arises when the semisoft cutoffs become replaced by quantities of $\order{\alpha_s^{1/2}\mu_B}$ and $\order{\mu_B}$. 

It is now clear that if we consider an expansion of $f$ about $P/Q = 0$
\begin{equation}
    f\!\left(\frac{P}{Q} , \Omega_{i}\right) = \cdots + a_{-1}(\Omega_{i}) \frac{Q}{P} + a_{0}(\Omega_{i}) + a_{1} (\Omega_{i}) \frac{P}{Q} + \cdots,
\end{equation}
the only term that will give a double logarithm will be the constant term $a_{0}$. This corresponds precisely to the $P \ll Q$ limit. Similarly, there is a contribution from $P \gg Q$, corresponding to an expansion of $f$ about ${Q / P = 0}$. Correctly accounting for the two integration regions reveals that the full double logarithm comes from the average of these contributions.

After extracting the average of the two series coefficients defined above, we are left with a double logarithm multiplying a (convergent) dimensionless angular integral given in eq.~(3) of the supplementary material, which can be computed analytically. The result is the coefficient $c_{3,2}$ of the $\alpha_{s}^{3} \ln^2 \alpha_{s}$ term in eq.~\eqref{eq:p_expansion}%
\footnote{The sign of our result in this equation is incorrect; the correct result is $-1$ times this expression. We thank Jean-Loïc Kneur for bringing this mistake to our attention.},
\begin{eqnarray}
    c_{3,2}\,\alpha_s^3 \ln^2 \alpha_s \!\!&=&\! - \frac{11}{48} \frac{N_c d_A}{(2 \pi)^3} \alpha_{s} m_\infty^4\ln^2 \alpha_{s}\nonumber\\
&=&\frac{3 (\mu_B/3)^4}{4\pi^2}\left[-0.266075 \, \alpha_s^3 \ln^2 \alpha_s\right],\,\,
\label{eq:main}
\end{eqnarray}
where the second equality holds for $N_c=N_f=3$. We have additionally verified that by repeating the calculation with $\Pi_T = m_\infty^2$ and $\Pi_{L} = 0$ from the outset, the result for $c_{3,2}$ remains unchanged, as was the case for the $\alpha_s^2 \ln \alpha_s$ term. Eq.~(\ref{eq:main}) is our main result. 
 
In order to elevate our result to the subleading-logarithm order $\order{\alpha_s^3 \ln \alpha_s}$, more care must be taken. Single logarithms may appear when only one of the loop momenta is semisoft while the other one is either soft or hard: If the other loop momentum is soft, a full HTL resummation of that line must be performed and the result cannot be expanded in powers of $\Pi_{T/L}$ as above. Meanwhile, if the other loop momentum is hard, no kinematic simplifications can be performed and no restrictions on topology and the number of fermion lines can be applied in that part of the diagram. In addition, the expansion of the soft one-loop diagram of eq.~\eqref{eq:ir_split} to higher orders in the soft loop momentum will lead to contributions of $\order{\alpha_s^3 \ln \alpha_s}$ that go beyond the HTL effective theory.

\emph{Conclusions}.---In the letter at hand, we have extracted the leading N$^3$LO correction to the pressure of cold quark matter using an existing two-loop computation within the Hard-Thermal-Loop effective theory. We note that the HTL result was derived in the different context of a hot quark-gluon plasma, but it is equally applicable to cold quark matter, as the soft contributions to the EoS are insensitive to the details of the physics at the hard scale ($T$ for a hot quark-gluon plasma and $\mu_B$ for cold quark matter). The hard scale appears in the calculation only through the asymptotic mass $m_\infty^2 \sim \alpha_{s} \int \dOne^3 \vec{p} f(\vec{p}) / |\vec{p}|$, where $f$ is the relevant distribution function.

\begin{figure}[t] 
\includegraphics[width=\linewidth]{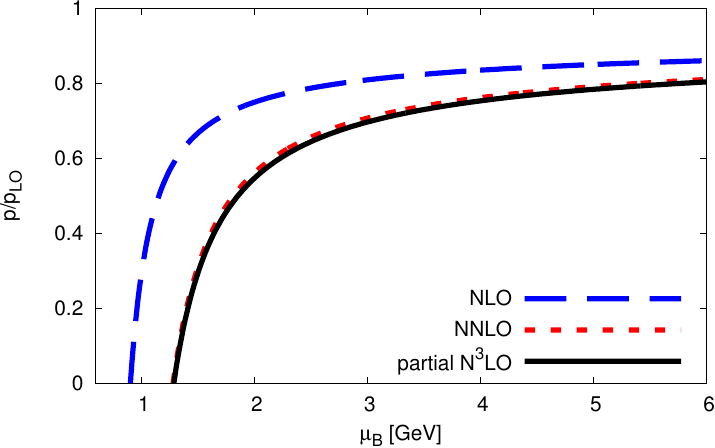}
\caption{\label{fig:pres_plot} The pressure of cold and dense massless QCD, normalized to the free pressure, as a function of baryon chemical potential for the renormalization scale choice $\bar\Lambda=2 \mu_B/3$ and $\Lambda_{\overline{\rm MS}}=0.378$~GeV.}
\end{figure}

We note that at higher orders, the semisoft contributions should continue to give rise to the leading logarithms $\alpha_{s}^{n + 1} \ln^{n} \alpha_{s}$. Quite strikingly, we find that the leading-logarithm contributions at NNLO and N$^{3}$LO are described by a theory with only two transverse gluons with a mass $m_\infty$. This leads us to conjecture that the leading-logarithm terms even at higher orders can be computed in this vastly simplified framework.

In Fig.~\ref{fig:pres_plot}, we display the pressure, evaluated with $\bar\Lambda=2 \mu_B/3$ and a two-loop running coupling, which indicates that the partial N$^{3}$LO term only constitutes a tiny correction to the existing NNLO result. One often estimates the error of a perturbative result such as eq.~\eqref{eq:p_expansion} by studying its dependence on the renormalization scale $\bar \Lambda$. However, the variation of this scale is completely insensitive to any as-yet-uncalculated soft physics: It is only sensitive to some subset of higher-order UV-sensitive terms in the weak-coupling series. As such, it is possible to grossly underestimate the systematic error by this procedure, as is the case at high $T$, where the soft contributions are even \emph{parametrically} larger than the next hard correction at any order (as they enter with odd powers of $\alpha_{s}^{1/2}$). That the leading-logarithm soft contribution at N$^{3}$LO gives a negligible correction to the NNLO result thus gives significant confidence in the error estimation of the previous results, and by extension increases confidence in using the perturbative result as \emph{ab-initio} input in calculations of the properties of neutron stars \cite{Kurkela:2009gj,Kurkela:2014vha,Gorda:2016uag,Annala:2017llu,Most:2018hfd} as well as simulations of gravitational-wave signals from neutron-star mergers.

By exploiting the same techniques that were outlined in the present work, we are confident that a calculation of the pressure to order $\order{\alpha_{s}^{3}\ln \alpha_{s}}$ is feasible, and plan to report the result of this calculation in the near future. 

\emph{Acknowledgments}.---TG, SS, and AV have been supported by the Academy of Finland grant no.~1303622, as well as by the European Research Council, grant no.~725369. PR is supported in part by the Department of Energy, DOE award no.~$\text{DE-SC0017905}$. We would like to thank Ioan Ghisoiu for his collaboration in the early stages of the project.

\bibliography{References.bib}

\newpage
\onecolumngrid
\input{appendix.txt}
\twocolumngrid

\end{document}

%% file: appendix.txt
\section{Supplemental Material: Details of Integration}

We present here some details of the calculation discussed in the main text. In particular, to carry out the four-momentum integrations such as $\int \dOne^4P$, we find it very useful to change variables from $(P^{0}, |\vec{p}|)$ to $(P, \Phi_{P})$, where $P$ is the magnitude of the Euclidean four-vector and $\Phi_{P}$ is the four-dimensional polar angle, ${\tan \Phi_{P} = |\vec{p}| / P^{0}}$. The particular expression that we use in these coordinates is only valid for ${0 \leq \Phi_{P} \leq \pi/2}$, but due to the symmetry of the self energy under $P^{0} \mapsto - P^{0}$, cf.~eqs.~(5) and (6) of the main text, one may use the measure 
\begin{align}
    \int_{\mathbb{R}^{4}} \!\dOne^{4} P &\mapsto \int_{0}^{\infty}\!  P^3 \dOne P \int_{0}^{\pi / 2}\!\! \dOne \Phi_{P} \,2 \sin^{2} \Phi_{P} \int_{S^{2}}\! \dOne^{2} \Omega_{P}
\end{align}
to restrict the integrations to this region.

As mentioned in the main text, the starting point in our N$^3$LO computation is the two-loop HTL pressure as written down in eq.~(34) of ref.~[27], where we convert the sum-integrals into ordinary 3+1 dimensional integrals because we work at $T = 0$. The full expression is rather unwieldy when written in full, and is not reproduced here, but we note that the simplifications outlined in the main text make extracting the double logarithm significantly easier.  An additional useful result is that, in the notation of ref.~[27], the propagator $\Delta_X = \order{\alpha_s}$, which allows us to discard a number of higher-order terms. 

When performing the expansions in $m_{\infty}^{2}$, we label the momenta using linear changes of variables $P \mapsto -P-Q$ and $Q \mapsto -P-Q$  when necessary, so that no factors of $\Pi_{T/L}(P + Q)$ appear. By dimensional analysis, this is always possible, as there are at most two propagators expanded beyond leading order that contribute to the $m_\infty^4$-term. This choice of momenta simplifies future integrations, as with it one only needs to consider the two regions $P \gg Q$ and $Q \gg P$ to obtain the double-logarithm coefficient. The full expression corresponding to eq.~(10) in the main text, that is, the two-loop HTL diagrams re-expanded to two self-energy insertions to obtain possible double-logarithm corrections at N$^3$LO, is
\begin{align}
&p_{\mathrm{IR,2}}^{\mathrm{HTL}}=-\frac{d_A N_c \alpha_s}{8\pi^5} \int_P \int_Q \int_{0}^{\pi/2} \int_{0}^{\pi/2} \int_{0}^{\pi} \frac{\mathrm{d}P}{P} \frac{\mathrm{d}Q}{Q}  \mathrm{d} \Phi_P \mathrm{d} \Phi_Q \mathrm{d} \theta \frac{\sin^3 \Phi_P \sin^3 \Phi_Q \sin \theta }{x^2\left[2x \cos\theta+\csc \Phi_P \csc \Phi_Q \left(x^2+1\right)+2x\cot\Phi_P\cot\Phi_Q\right]} \times \nonumber \\
&\times \Bigg \lbrace 4x m_\infty^2 \pilp \csc^6 \Phi_P \csc \Phi_Q \left[ x \csc\Phi_Q \left(1-3\Phi_P\cot\Phi_P + 2 \Phi_Q \cot \Phi_Q\right)- 2 \csc \Phi_P \cos\theta \left(\Phi_P \cot\Phi_P -1 \right)\right] \nonumber \\
&\phantom{\times \lbrace}+ 4 xm_\infty^2 \pitp \csc \Phi_Q \csc \Phi_P \left[ 2\csc\Phi_P \cos\theta\left(\Phi_P \cot\Phi_P -1\right)+x\csc\Phi_Q \left(3\Phi_P\cot\Phi_P-2\Phi_Q\cot\Phi_Q-1\right) \right]  \nonumber \\
&\phantom{\times \lbrace}+ x^2 \pilp \pilq \csc^3 \Phi_P \csc^2 \Phi_Q  \Big[ - 4 x \csc^2 \Phi_P \csc\Phi_Q \cos \theta + 2 x \csc\Phi_Q \cos \theta + 3 x^2 \csc \Phi_P \csc^4 \Phi_Q \nonumber \\
&\phantom{\times \lbrace}\quad-\csc^3\Phi_P \csc^2\Phi_Q+4 x  \cot \Phi_P \csc^2 \Phi_P \cot\Phi_Q \csc \Phi_Q  + 2x \cot \Phi_P \cot\Phi_Q \csc \Phi_Q  + 4 \csc^3 \Phi_P \cot^2 \Phi_Q \Big] \nonumber \\ 
&\phantom{\times \lbrace}+ x^2 \pitp \pitq \csc \Phi_Q \csc \Phi_P \Big[ -4 x\csc^2\Phi_P \left(x-\cot\Phi_P \cot\Phi_Q\right) + 4 \cos\theta x \left(3\csc^2\Phi_Q -2 \right) \nonumber\\
&\phantom{\times \lbrace}\quad+\csc\Phi_P\csc^3\Phi_Q \left(-x^2 \cos\left(2\Phi_Q\right)-5x^2\cos\left(2\Phi_P\right)+3x^2+6\right) + 3x^2 \cot^2\Phi_P\csc\Phi_P\csc\Phi_Q \nonumber \\
&\phantom{\times \lbrace}\quad-9\cot^2\Phi_Q \csc\Phi_P \csc\Phi_Q+ 2x\cot\Phi_P\csc^3\Phi_Q\left(\cos\left(3\Phi_Q\right)-7\cos\Phi_Q\right) - 3\csc^3\Phi_P \csc\Phi_Q \left(2\cos\left(2\Phi_Q\right)+2\right) \Big]\nonumber \\
&\phantom{\times \lbrace}+ x^2 \pilp \pitq \csc^3\Phi_P \csc\Phi_Q \Big[x \csc^2\Phi_P \left(4 \cos \theta + \cot\Phi_P \cot \Phi_Q \cos\left(2\Phi_P\right) - 5 \cot\Phi_P \cot\Phi_Q\right) - 2 x\cos\theta  \nonumber \\
&\phantom{\times \lbrace}\quad+ x^2  \csc\Phi_P \csc\Phi_Q \left(4-3\cot^2\Phi_P - 3 \csc^2 \Phi_Q \right) + 3x \csc^3 \Phi_P \csc\Phi_Q \left(\cos\left(2\Phi_Q\right)+2\right)  \Big ] \nonumber \\
&\phantom{\times \lbrace}+ x^2 \pitp \pilq \csc^2 \Phi_Q  \Big [ \csc^4 \Phi_P \csc^2 \Phi_Q - x\cos\theta\csc\Phi_P \csc^3\Phi_Q \left(\cos\left(2\Phi_Q\right)+11\right) \nonumber \\
&\phantom{\times \lbrace}\quad +4x\cos\theta\csc^3 \Phi_P\csc\Phi_Q  +x^2 \csc^4 \Phi_Q \left(3\cot^2\Phi_P-7\right) + \cot^2 \Phi_Q \left( 3 \csc^2\Phi_P\csc^2\Phi_Q - 4 \csc^4 \Phi_P \right)  \nonumber \\
&\phantom{\times \lbrace}\quad  + 2x\cot\Phi_P \csc\Phi_P \cot\Phi_Q \csc\Phi_Q \left(1-2\csc^2\Phi_P + 6\csc^2 \Phi_Q \right) \Big] \nonumber \\
&\phantom{\times \lbrace}- \frac{1}{2} \Pi_L^2 \left(\Phi_P\right) \csc^8 \Phi_P \csc^2 \Phi_Q \Big [ 32x \cos\theta \sin\Phi_P \sin\Phi_Q + 28 x \cos\Phi_P \cos\Phi_Q \nonumber \\
&\phantom{\times \lbrace}\quad -5x^2 \sin^2 \Phi_P+5x^2 \cos^2 \Phi_P + 8 \sin^2 \Phi_Q - 8 \cos^2 \Phi_Q + 13 x^2 \Big ] \nonumber \\
&\phantom{\times \lbrace}+ \Pi_T^2 \left(\Phi_P\right) \csc\Phi_Q \csc\Phi_P \Big [ 16 x\cos \theta \csc^2 \Phi_P -14 x\cos \theta + 14 x \cot\Phi_P^3 \cot\Phi_Q \nonumber \\
&\phantom{\times \lbrace}\quad + \csc^3 \Phi_P \csc\Phi_Q \left( x^2 \cos\left(2\Phi_P\right)-4\cos\left(2\Phi_P\right) - 4 \cos\left(2\Phi_Q \right) +8x^2 -4 \right) \Big ] \Bigg \rbrace,
\label{eq:fullintegral}
\end{align}
where $x\equiv P/Q$ and $\theta$ is the angle between $\vec{p}$ and $\vec{q}$. 
After expanding in the limits $P \gg Q$, $Q \gg P$ and averaging, we obtain the pure double-logarithm N$^3$LO contribution in the following form:
\begin{align}
& p^{\mathrm{semisoft}}_{\mathrm{IR},2}=\frac{d_A N_c \alpha_s m_\infty^4}{256\pi^5} 
\int_{\Lambda_1^P}^{\Lambda_2^P} \int_{\Lambda_1^Q}^{\Lambda_2^Q}  \int_{0}^{\pi/2} \int_{0}^{\pi/2} \int_{0}^{\pi} \frac{\mathrm{d}P}{P}\frac{\mathrm{d}Q}{Q} \mathrm{d} \Phi_P \mathrm{d} \Phi_Q \mathrm{d} \theta
\sin^2 \Phi_P \sin^2 \Phi_Q \sin \theta
\Bigg \lbrace \frac{1}{16}\csc^8 \Phi_P \cos \left(2\Phi_Q\right) \times \nonumber \\
&\times \Big[ 2\left(\left(-16\Phi_P^2+677\right)\cos^2\theta-2384\Phi_P^2-615\right) 
+2\cos\left(2\Phi_P\right)\left( \left( 8\Phi_P^2-991 \right)\cos^2\theta-3448\Phi_P^2 -555\right) \nonumber \\
&\quad+8\cos\left(4\Phi_P\right) \left( \left(4\Phi_P^2+89\right)\cos^2\theta-268\Phi_P^2+319\right)
-\cos\left(6\Phi_P\right) \left( \left(16\Phi_P^2+67\right)\cos^2\theta+16\Phi_P^2+171\right) \nonumber \\
&\quad-6\cos\left(8\Phi_P\right)\left(3\cos^2\theta+7\right)
+\cos\left(10\Phi_P\right)\left(\cos^2\theta+1\right)
+ 16\Phi_P \sin\left(2\Phi_P\right) \left(403-57\cos^2\theta\right) \nonumber \\
&\quad+16\Phi_P\sin\left(4\Phi_P\right) \left(47\cos^2\theta+261\right)
+ 16 \Phi_P \sin\left(6\Phi_P\right) \left(13\cos^2\theta+21\right)
+ 8\Phi_P\sin\left(8\Phi_P\right)\left(\cos^2\theta+1\right) \Big]\nonumber \\
&+16\Big[ \left(4\Phi_P\cot\Phi_P-\cos\left(2\Phi_P\right)+9\right)\cos^2\theta
-4\Phi_P\cot\Phi_P+\cos\left(2\Phi_P\right)-27 \nonumber \\
&\quad+2\csc^2\Phi_P \left( \left(14\Phi_P\cot\Phi_P-\Phi_P^2-15\right)\cos^2\theta
-23\Phi_P\cot\Phi_P+\Phi_P^2+25 \right) \nonumber \\
&\quad+\csc^4\Phi_P\big( 2\left(12\Phi_P\cot\Phi_P+7\Phi_P^2-6 \right) \cos^2\theta
+9\Phi_P\cot\Phi_P+4\Phi_P^2-3 \big) \nonumber \\ 
&\quad+3\csc^6\Phi_P\left(18\Phi_P\cot\Phi_P+ \left(7-4\cos^2\theta\right)\Phi_P^2-9\right)
-27\Phi_P^2\csc^8 \Phi_P \Big] \nonumber \\
&+\frac{1}{2} \Big[-1040\Phi_P^2-1569
+22\cos\left(2\Phi_P\right)\left(77-48\Phi_P^2\right)
-16\cos\left(4\Phi_P\right)\left(\Phi_P^2+6\right)
-30\cos\left(6\Phi_P\right)
+\cos\left(8\Phi_P\right) \nonumber \\
&\quad + 8\Phi_P\left(329\sin\left(2\Phi_P\right)
-34\sin\left(4\Phi_P\right)
+\sin\left(6\Phi_P\right)\right) \Big]
 \cos\theta\cot\Phi_P\csc^6\Phi_P\sin\left(2\Phi_Q\right) \nonumber \\
&-\csc^4\Phi_P\cos\left(4\Phi_Q\right)\left(\cos^2\theta-\cot^2\Phi_P\right) 
\left[\cos\left(2\Phi_P\right)-6\Phi_P\cot\Phi_P+5\right]
\left[\cos\left(2\Phi_P\right)+2\Phi_P\cot\Phi_P-3\right] \nonumber \\
&-8\cos\theta\csc^6\Phi_P\sin\left(4\Phi_Q\right)
\left[4\Phi_P\cos\Phi_P-7\sin\Phi_P+\sin\left(3\Phi_P\right)\right]
\left[12\Phi_P\cos\Phi_P-9\sin\Phi_P-\sin\left(3\Phi_P\right)\right] \nonumber \\
&+48 \left(\Phi_Q\cot\Phi_Q-1\right) \csc^4\Phi_Q 
\left[5\cos\left(2\Phi_P\right)-10\Phi_P\cot\Phi_P+11\right]
+80 \csc^2\Phi_Q \left[\cos\left(2\Phi_P\right)-2\Phi_P\cot\Phi_P+5\right]  \nonumber \\
&+40 \Phi_Q\csc^5\Phi_P\cot\Phi_Q
 \left[12\Phi_P\cos\Phi_p-9\sin\Phi_P-\sin\left(3\Phi_P\right) \right] 
-224 \Phi_Q\cot\Phi_Q\csc^2\Phi_Q \Bigg\rbrace.
\label{eq:angularintegral}
\end{align}
 The integral over the three-dimensional angle $\theta$ can then be performed analytically, which yields a more manageable expression, 
\begin{align}
&p^{\mathrm{semisoft}}_{\mathrm{IR},2}= \frac{d_A N_c \alpha_s m_\infty^4}{256\pi^5} 
\int_{\Lambda_1^P}^{\Lambda_2^P} \int_{\Lambda_1^Q}^{\Lambda_2^Q} \int_{0}^{\pi/2} \int_{0}^{\pi/2}\frac{\mathrm{d}P}{P}\frac{\mathrm{d}Q}{Q} \mathrm{d} \Phi_P \mathrm{d} \Phi_Q 
\sin^2 \Phi_P \sin^2 \Phi_Q 
\Bigg \lbrace \frac{1}{6}\csc^8 \Phi_P \cos \left(2\Phi_Q\right) \times \nonumber \\
&\times\Big[ -16\cos\left(2\Phi_P\right)\left(323\Phi_P^2+83\right)
-4\cos\left(4\Phi_P\right)\left(400\Phi_P^2-523\right)
-\cos\left(6\Phi_P\right)\left(2\Phi_P+145\right)
-36\cos\left(8\Phi_P\right)+\cos\left(10\Phi_P\right) \nonumber \\
&\quad + 8\left(576\sin\left(2\Phi_P\right)
+415\sin\left(4\Phi_P\right)
-38\sin\left(6\Phi_P\right)
+8\sin\left(8\Phi_P\right)\right)\Phi_P
-3584\Phi_P^2-584 \Big] \nonumber \\
&+\frac{32}{3} \Big[\left(-8\Phi_P\cot\Phi_P+2\cos\left(2\Phi_P\right)-72\right)
-2\csc^2\Phi_P\left(55\Phi_P\cot\Phi_P-2\Phi_P^2-60\right)\nonumber \\
&\quad+\csc^4\Phi_P\left(51\Phi_P\cot\Phi_P+26\Phi_P^2-21\right) +3\csc^6\Phi_P\left(54\Phi_P\cot\Phi_P+17\Phi_P^2-27\right)
-81\csc^8\Phi_P\Phi_P^2 \Big] \nonumber \\
&-160\csc^2\Phi_Q \left[2\Phi_P\cot\Phi_P-\cos\left(2\Phi_P\right)-5\right] 
-96\left(\Phi_Q\cot\Phi_Q-1\right) \csc^4 \Phi_Q
\left[ 10\Phi_P\cot\Phi_P- 5\cos\left(2\Phi_P\right)-11\right]
\nonumber \\
&+\frac{8}{3} \cos\left(4\Phi_Q\right)\csc^8\Phi_P 
\left[2\cos\left(2\Phi_P\right)+1\right]
\left[4\Phi_P\cos\Phi_P-7\sin\Phi_P+\sin\left(3\Phi_P\right)\right] \nonumber \\
& +80 \Phi_Q\cot\Phi_Q\csc^5\Phi_P
\left[12\Phi_P\cos\Phi_P-9\sin\Phi_P-\sin\left(3\Phi_P\right)\right] 
-448\Phi_Q\cot\Phi_Q\csc^2\Phi_Q \Bigg \rbrace.
\label{eq:4dintegral}
\end{align}
 The last two angular integrals are likewise easy to perform, and together with the $\log^2 \alpha_s^{1/2}$ from the radial integrals, this yields precisely the promised coefficient of eq.~(13).

For comparison, we may consider the simplification of setting $\Pi_T = m_\infty^2$, $\Pi_L = 0$.  The integrals corresponding to eqs.~(15)-(17) are then 
  \begin{align}
&p_{\mathrm{IR,2}}^{m^{2}_{\infty}}=-\frac{d_A N_c \alpha_s}{8\pi^5} \int_P \int_Q \int_{0}^{\pi/2} \int_{0}^{\pi/2} \int_{0}^{\pi} \frac{\mathrm{d}P}{P} \frac{\mathrm{d}Q}{Q}  \mathrm{d} \Phi_P \mathrm{d} \Phi_Q \mathrm{d} \theta \frac{m_\infty^4 \sin \theta}{x^2\left[2x \cos\theta\sin \Phi_P \sin \Phi_Q+2x\cos\Phi_P\cos\Phi_Q + x^2+1\right]} \times \nonumber \\
&\quad \times \Bigg \lbrace 4x \cos\theta \sin\Phi_P \sin^3\Phi_Q \left[x^2-4\right]-x\cos\theta\sin^3\Phi_P\sin\Phi_Q \left[\left(4x^2+7\right)\cos\left(2\Phi_Q\right)+8x^2-7\right] \nonumber \\
&\phantom{\lbrace}\quad\quad+\frac{x\cos\theta\sin^3\Phi_P\sin^3\Phi_Q}{2}\left[4x\cos\Phi_P\cos\Phi_Q\left(2x^2+7\right)+3x^2\left(x^2+1\right)\left(\cos\left(2\Phi_P\right)+1\right)-2\left(3x^2-4\right)\cos\left(2\Phi_Q\right)\right]\nonumber \\
&\phantom{\lbrace}\quad\quad +2x \cos\Phi_P\cos\Phi_Q\sin^2\Phi_P\left[\sin^2\Phi_Q\left(4x^2+7\right)+6x^2\right]+\sin^2\Phi_P\cos\left(2\Phi_Q\right)\left[x^4+5x^2-4\right] \nonumber \\
&\phantom{\lbrace}\quad\quad - 2\sin^2\Phi_P \left[5x^4+x^2-1\right] + 2x^4\sin^4\Phi_P +3x^4 \sin^2 \left(2\Phi_P\right) \Bigg \rbrace,
\end{align}
 \begin{align}
&p^{\mathrm{semisoft}}_{\mathrm{IR},2}= \frac{d_A N_c \alpha_s m_\infty^4}{32\pi^5}
\int_{\Lambda_1^P}^{\Lambda_2^P} \int_{\Lambda_1^Q}^{\Lambda_2^Q} \int_{0}^{\pi/2} \int_{0}^{\pi/2} \int_{0}^{\pi}\frac{\mathrm{d}P}{P}\frac{\mathrm{d}Q}{Q}  \mathrm{d} \Phi_P \mathrm{d} \Phi_Q \mathrm{d} \theta
\sin^2 \Phi_P \sin^2 \Phi_Q \sin \theta
\Bigg \lbrace -\frac{1}{2}\csc^2 \Phi_P \cos \left(2\Phi_Q\right) \times \nonumber \\
& \quad\times \left[ 19\cos^2\theta-41-4\cos\left(2\Phi_P\right)\left(5\cos^2\theta+11\right)+\cos\left(4\Phi_P\right)\left(\cos^2\theta+1\right) \right] \nonumber \\
& \quad +26\cos^2\theta-62- 2\cos\left(2\Phi_P\right)\left[\cos^2\theta-1\right]+8\cos\left(4\Phi_Q\right)\left[\cot^2\Phi_P-\cos^2\theta\right]+16\sin\left(4\Phi_Q\right)\cot\Phi_P\cos\theta \nonumber \\
&\quad -4\sin\left(2\Phi_Q\right)\cot\Phi_P\cos\theta\left[\cos\left(2\Phi_P\right)-16\right]+24\csc^2\Phi_P+2\csc^2\Phi_Q\left[5\cos\left(2\Phi_P\right)+1\right] \Bigg \rbrace
 \end{align}
and 

\begin{align}
&p^{\mathrm{semisoft}}_{\mathrm{IR},2}= \frac{d_A N_c \alpha_s m_\infty^4}{24\pi^5} 
\int_{\Lambda_1^P}^{\Lambda_2^P} \int_{\Lambda_1^Q}^{\Lambda_2^Q} \int_{0}^{\pi/2} \int_{0}^{\pi/2}\frac{\mathrm{d}P}{P}\frac{\mathrm{d}Q}{Q} \mathrm{d} \Phi_P \mathrm{d} \Phi_Q 
\sin^2 \Phi_P \sin^2 \Phi_Q \times \nonumber \\
&\quad\times\Bigg \lbrace -80+ \cos\left(2\Phi_Q\right)\csc^2\Phi_P \left[38\cos\left(2\Phi_P\right)-4\cos\left(4\Phi_P\right) \right] +2\cos\left(2\Phi_P\right) \nonumber \\
&\phantom{\times \lbrace}\quad+4\cos\left(4\Phi_Q\right)\csc^2\Phi_P\left[2\cos\left(2\Phi_P\right)+1\right]+3\csc^2\Phi_Q\left[5\cos\left(2\Phi_P\right)+1\right]+36\csc^2\Phi_P \Bigg \rbrace, 
\end{align}
which evidently represent a significant simplification  in comparison with the full expressions, yet lead to the same results. 

\clearpage

%% file: dense_pqcd_PRL.bbl
\begin{thebibliography}{32}%
\makeatletter
\providecommand \@ifxundefined [1]{%
 \@ifx{#1\undefined}
}%
\providecommand \@ifnum [1]{%
 \ifnum #1\expandafter \@firstoftwo
 \else \expandafter \@secondoftwo
 \fi
}%
\providecommand \@ifx [1]{%
 \ifx #1\expandafter \@firstoftwo
 \else \expandafter \@secondoftwo
 \fi
}%
\providecommand \natexlab [1]{#1}%
\providecommand \enquote  [1]{``#1''}%
\providecommand \bibnamefont  [1]{#1}%
\providecommand \bibfnamefont [1]{#1}%
\providecommand \citenamefont [1]{#1}%
\providecommand \href@noop [0]{\@secondoftwo}%
\providecommand \href [0]{\begingroup \@sanitize@url \@href}%
\providecommand \@href[1]{\@@startlink{#1}\@@href}%
\providecommand \@@href[1]{\endgroup#1\@@endlink}%
\providecommand \@sanitize@url [0]{\catcode `\\12\catcode `\$12\catcode `\&12\catcode `\#12\catcode `\^12\catcode `\_12\catcode `\%12\relax}%
\providecommand \@@startlink[1]{}%
\providecommand \@@endlink[0]{}%
\providecommand \url  [0]{\begingroup\@sanitize@url \@url }%
\providecommand \@url [1]{\endgroup\@href {#1}{\urlprefix }}%
\providecommand \urlprefix  [0]{URL }%
\providecommand \Eprint [0]{\href }%
\providecommand \doibase [0]{http://dx.doi.org/}%
\providecommand \selectlanguage [0]{\@gobble}%
\providecommand \bibinfo  [0]{\@secondoftwo}%
\providecommand \bibfield  [0]{\@secondoftwo}%
\providecommand \translation [1]{[#1]}%
\providecommand \BibitemOpen [0]{}%
\providecommand \bibitemStop [0]{}%
\providecommand \bibitemNoStop [0]{.\EOS\space}%
\providecommand \EOS [0]{\spacefactor3000\relax}%
\providecommand \BibitemShut  [1]{\csname bibitem#1\endcsname}%
\let\auto@bib@innerbib\@empty
%</preamble>
\bibitem [{\citenamefont {Aarts}(2012)}]{Aarts:2013lcm}%
  \BibitemOpen
  \bibfield  {author} {\bibinfo {author} {\bibfnamefont {G.}~\bibnamefont {Aarts}},\ }\bibfield  {booktitle} {\emph {\bibinfo {booktitle} {{Proceedings, 30th International Symposium on Lattice Field Theory (Lattice 2012)}}},\ }\href@noop {} {\bibfield  {journal} {\bibinfo  {journal} {PoS}\ }\textbf {\bibinfo {volume} {LATTICE2012}},\ \bibinfo {pages} {017} (\bibinfo {year} {2012})},\ \Eprint {http://arxiv.org/abs/1302.3028} {arXiv:1302.3028 [hep-lat]} \BibitemShut {NoStop}%
%%CITATION = ARXIV:1302.3028;%%
\bibitem [{\citenamefont {Fodor}\ \emph {et~al.}(2015)\citenamefont {Fodor}, \citenamefont {Katz}, \citenamefont {Sexty},\ and\ \citenamefont {Török}}]{Fodor:2015doa}%
  \BibitemOpen
  \bibfield  {author} {\bibinfo {author} {\bibfnamefont {Z.}~\bibnamefont {Fodor}}, \bibinfo {author} {\bibfnamefont {S.~D.}\ \bibnamefont {Katz}}, \bibinfo {author} {\bibfnamefont {D.}~\bibnamefont {Sexty}}, \ and\ \bibinfo {author} {\bibfnamefont {C.}~\bibnamefont {Török}},\ }\href@noop {} {\  (\bibinfo {year} {2015})},\ \Eprint {http://arxiv.org/abs/1508.05260} {arXiv:1508.05260 [hep-lat]} \BibitemShut {NoStop}%
%%CITATION = ARXIV:1508.05260;%%
\bibitem [{\citenamefont {Cristoforetti}\ \emph {et~al.}(2012)\citenamefont {Cristoforetti}, \citenamefont {Di~Renzo},\ and\ \citenamefont {Scorzato}}]{Cristoforetti:2012su}%
  \BibitemOpen
  \bibfield  {author} {\bibinfo {author} {\bibfnamefont {M.}~\bibnamefont {Cristoforetti}}, \bibinfo {author} {\bibfnamefont {F.}~\bibnamefont {Di~Renzo}}, \ and\ \bibinfo {author} {\bibfnamefont {L.}~\bibnamefont {Scorzato}} (\bibinfo {collaboration} {AuroraScience}),\ }\href {\doibase 10.1103/PhysRevD.86.074506} {\bibfield  {journal} {\bibinfo  {journal} {Phys. Rev.}\ }\textbf {\bibinfo {volume} {D86}},\ \bibinfo {pages} {074506} (\bibinfo {year} {2012})},\ \Eprint {http://arxiv.org/abs/1205.3996} {arXiv:1205.3996 [hep-lat]} \BibitemShut {NoStop}%
%%CITATION = ARXIV:1205.3996;%%
\bibitem [{\citenamefont {Fujii}\ \emph {et~al.}(2013)\citenamefont {Fujii}, \citenamefont {Honda}, \citenamefont {Kato}, \citenamefont {Kikukawa}, \citenamefont {Komatsu},\ and\ \citenamefont {Sano}}]{Fujii:2013sra}%
  \BibitemOpen
  \bibfield  {author} {\bibinfo {author} {\bibfnamefont {H.}~\bibnamefont {Fujii}}, \bibinfo {author} {\bibfnamefont {D.}~\bibnamefont {Honda}}, \bibinfo {author} {\bibfnamefont {M.}~\bibnamefont {Kato}}, \bibinfo {author} {\bibfnamefont {Y.}~\bibnamefont {Kikukawa}}, \bibinfo {author} {\bibfnamefont {S.}~\bibnamefont {Komatsu}}, \ and\ \bibinfo {author} {\bibfnamefont {T.}~\bibnamefont {Sano}},\ }\href {\doibase 10.1007/JHEP10(2013)147} {\bibfield  {journal} {\bibinfo  {journal} {JHEP}\ }\textbf {\bibinfo {volume} {10}},\ \bibinfo {pages} {147} (\bibinfo {year} {2013})},\ \Eprint {http://arxiv.org/abs/1309.4371} {arXiv:1309.4371 [hep-lat]} \BibitemShut {NoStop}%
%%CITATION = ARXIV:1309.4371;%%
\bibitem [{\citenamefont {de~Forcrand}\ \emph {et~al.}(2014)\citenamefont {de~Forcrand}, \citenamefont {Langelage}, \citenamefont {Philipsen},\ and\ \citenamefont {Unger}}]{deForcrand:2014tha}%
  \BibitemOpen
  \bibfield  {author} {\bibinfo {author} {\bibfnamefont {P.}~\bibnamefont {de~Forcrand}}, \bibinfo {author} {\bibfnamefont {J.}~\bibnamefont {Langelage}}, \bibinfo {author} {\bibfnamefont {O.}~\bibnamefont {Philipsen}}, \ and\ \bibinfo {author} {\bibfnamefont {W.}~\bibnamefont {Unger}},\ }\href {\doibase 10.1103/PhysRevLett.113.152002} {\bibfield  {journal} {\bibinfo  {journal} {Phys. Rev. Lett.}\ }\textbf {\bibinfo {volume} {113}},\ \bibinfo {pages} {152002} (\bibinfo {year} {2014})},\ \Eprint {http://arxiv.org/abs/1406.4397} {arXiv:1406.4397 [hep-lat]} \BibitemShut {NoStop}%
%%CITATION = ARXIV:1406.4397;%%
\bibitem [{\citenamefont {Ichihara}\ \emph {et~al.}(2015)\citenamefont {Ichihara}, \citenamefont {Morita},\ and\ \citenamefont {Ohnishi}}]{Ichihara:2015kba}%
  \BibitemOpen
  \bibfield  {author} {\bibinfo {author} {\bibfnamefont {T.}~\bibnamefont {Ichihara}}, \bibinfo {author} {\bibfnamefont {K.}~\bibnamefont {Morita}}, \ and\ \bibinfo {author} {\bibfnamefont {A.}~\bibnamefont {Ohnishi}},\ }\href@noop {} {\  (\bibinfo {year} {2015})},\ \Eprint {http://arxiv.org/abs/1507.04527} {arXiv:1507.04527 [hep-lat]} \BibitemShut {NoStop}%
%%CITATION = ARXIV:1507.04527;%%
\bibitem [{\citenamefont {Alexandru}\ \emph {et~al.}(2016)\citenamefont {Alexandru}, \citenamefont {Basar}, \citenamefont {Bedaque}, \citenamefont {Ridgway},\ and\ \citenamefont {Warrington}}]{Alexandru:2015sua}%
  \BibitemOpen
  \bibfield  {author} {\bibinfo {author} {\bibfnamefont {A.}~\bibnamefont {Alexandru}}, \bibinfo {author} {\bibfnamefont {G.}~\bibnamefont {Basar}}, \bibinfo {author} {\bibfnamefont {P.~F.}\ \bibnamefont {Bedaque}}, \bibinfo {author} {\bibfnamefont {G.~W.}\ \bibnamefont {Ridgway}}, \ and\ \bibinfo {author} {\bibfnamefont {N.~C.}\ \bibnamefont {Warrington}},\ }\href {\doibase 10.1007/JHEP05(2016)053} {\bibfield  {journal} {\bibinfo  {journal} {JHEP}\ }\textbf {\bibinfo {volume} {05}},\ \bibinfo {pages} {053} (\bibinfo {year} {2016})},\ \Eprint {http://arxiv.org/abs/1512.08764} {arXiv:1512.08764 [hep-lat]} \BibitemShut {NoStop}%
%%CITATION = ARXIV:1512.08764;%%
\bibitem [{\citenamefont {Gross}\ and\ \citenamefont {Wilczek}(1973)}]{PhysRevLett.30.1343}%
  \BibitemOpen
  \bibfield  {author} {\bibinfo {author} {\bibfnamefont {D.~J.}\ \bibnamefont {Gross}}\ and\ \bibinfo {author} {\bibfnamefont {F.}~\bibnamefont {Wilczek}},\ }\href {\doibase 10.1103/PhysRevLett.30.1343} {\bibfield  {journal} {\bibinfo  {journal} {Phys. Rev. Lett.}\ }\textbf {\bibinfo {volume} {30}},\ \bibinfo {pages} {1343} (\bibinfo {year} {1973})}\BibitemShut {NoStop}%
\bibitem [{\citenamefont {Freedman}\ and\ \citenamefont {McLerran}(1977{\natexlab{a}})}]{Freedman:1976dm}%
  \BibitemOpen
  \bibfield  {author} {\bibinfo {author} {\bibfnamefont {B.~A.}\ \bibnamefont {Freedman}}\ and\ \bibinfo {author} {\bibfnamefont {L.~D.}\ \bibnamefont {McLerran}},\ }\href {\doibase 10.1103/PhysRevD.16.1147} {\bibfield  {journal} {\bibinfo  {journal} {Phys. Rev.}\ }\textbf {\bibinfo {volume} {D16}},\ \bibinfo {pages} {1147} (\bibinfo {year} {1977}{\natexlab{a}})}\BibitemShut {NoStop}%
%%CITATION = PHRVA,D16,1147;%%
\bibitem [{\citenamefont {Freedman}\ and\ \citenamefont {McLerran}(1977{\natexlab{b}})}]{Freedman:1976ub}%
  \BibitemOpen
  \bibfield  {author} {\bibinfo {author} {\bibfnamefont {B.~A.}\ \bibnamefont {Freedman}}\ and\ \bibinfo {author} {\bibfnamefont {L.~D.}\ \bibnamefont {McLerran}},\ }\href {\doibase 10.1103/PhysRevD.16.1169} {\bibfield  {journal} {\bibinfo  {journal} {Phys. Rev.}\ }\textbf {\bibinfo {volume} {D16}},\ \bibinfo {pages} {1169} (\bibinfo {year} {1977}{\natexlab{b}})}\BibitemShut {NoStop}%
%%CITATION = PHRVA,D16,1169;%%
\bibitem [{\citenamefont {Vuorinen}(2003)}]{Vuorinen:2003fs}%
  \BibitemOpen
  \bibfield  {author} {\bibinfo {author} {\bibfnamefont {A.}~\bibnamefont {Vuorinen}},\ }\href {\doibase 10.1103/PhysRevD.68.054017} {\bibfield  {journal} {\bibinfo  {journal} {Phys. Rev.}\ }\textbf {\bibinfo {volume} {D68}},\ \bibinfo {pages} {054017} (\bibinfo {year} {2003})},\ \Eprint {http://arxiv.org/abs/hep-ph/0305183} {arXiv:hep-ph/0305183 [hep-ph]} \BibitemShut {NoStop}%
%%CITATION = HEP-PH/0305183;%%
\bibitem [{\citenamefont {Ipp}\ \emph {et~al.}(2006)\citenamefont {Ipp}, \citenamefont {Kajantie}, \citenamefont {Rebhan},\ and\ \citenamefont {Vuorinen}}]{Ipp:2006ij}%
  \BibitemOpen
  \bibfield  {author} {\bibinfo {author} {\bibfnamefont {A.}~\bibnamefont {Ipp}}, \bibinfo {author} {\bibfnamefont {K.}~\bibnamefont {Kajantie}}, \bibinfo {author} {\bibfnamefont {A.}~\bibnamefont {Rebhan}}, \ and\ \bibinfo {author} {\bibfnamefont {A.}~\bibnamefont {Vuorinen}},\ }\href {\doibase 10.1103/PhysRevD.74.045016} {\bibfield  {journal} {\bibinfo  {journal} {Phys. Rev.}\ }\textbf {\bibinfo {volume} {D74}},\ \bibinfo {pages} {045016} (\bibinfo {year} {2006})},\ \Eprint {http://arxiv.org/abs/hep-ph/0604060} {arXiv:hep-ph/0604060 [hep-ph]} \BibitemShut {NoStop}%
%%CITATION = HEP-PH/0604060;%%
\bibitem [{\citenamefont {Kurkela}\ and\ \citenamefont {Vuorinen}(2016)}]{Kurkela:2016was}%
  \BibitemOpen
  \bibfield  {author} {\bibinfo {author} {\bibfnamefont {A.}~\bibnamefont {Kurkela}}\ and\ \bibinfo {author} {\bibfnamefont {A.}~\bibnamefont {Vuorinen}},\ }\href {\doibase 10.1103/PhysRevLett.117.042501} {\bibfield  {journal} {\bibinfo  {journal} {Phys. Rev. Lett.}\ }\textbf {\bibinfo {volume} {117}},\ \bibinfo {pages} {042501} (\bibinfo {year} {2016})},\ \Eprint {http://arxiv.org/abs/1603.00750} {arXiv:1603.00750 [hep-ph]} \BibitemShut {NoStop}%
%%CITATION = ARXIV:1603.00750;%%
\bibitem [{\citenamefont {Fraga}\ and\ \citenamefont {Romatschke}(2005)}]{Fraga:2004gz}%
  \BibitemOpen
  \bibfield  {author} {\bibinfo {author} {\bibfnamefont {E.~S.}\ \bibnamefont {Fraga}}\ and\ \bibinfo {author} {\bibfnamefont {P.}~\bibnamefont {Romatschke}},\ }\href {\doibase 10.1103/PhysRevD.71.105014} {\bibfield  {journal} {\bibinfo  {journal} {Phys. Rev.}\ }\textbf {\bibinfo {volume} {D71}},\ \bibinfo {pages} {105014} (\bibinfo {year} {2005})},\ \Eprint {http://arxiv.org/abs/hep-ph/0412298} {arXiv:hep-ph/0412298 [hep-ph]} \BibitemShut {NoStop}%
%%CITATION = HEP-PH/0412298;%%
\bibitem [{\citenamefont {Kurkela}\ \emph {et~al.}(2010)\citenamefont {Kurkela}, \citenamefont {Romatschke},\ and\ \citenamefont {Vuorinen}}]{Kurkela:2009gj}%
  \BibitemOpen
  \bibfield  {author} {\bibinfo {author} {\bibfnamefont {A.}~\bibnamefont {Kurkela}}, \bibinfo {author} {\bibfnamefont {P.}~\bibnamefont {Romatschke}}, \ and\ \bibinfo {author} {\bibfnamefont {A.}~\bibnamefont {Vuorinen}},\ }\href {\doibase 10.1103/PhysRevD.81.105021} {\bibfield  {journal} {\bibinfo  {journal} {Phys. Rev.}\ }\textbf {\bibinfo {volume} {D81}},\ \bibinfo {pages} {105021} (\bibinfo {year} {2010})},\ \Eprint {http://arxiv.org/abs/0912.1856} {arXiv:0912.1856 [hep-ph]} \BibitemShut {NoStop}%
%%CITATION = ARXIV:0912.1856;%%
\bibitem [{\citenamefont {Fraga}\ \emph {et~al.}(2014)\citenamefont {Fraga}, \citenamefont {Kurkela},\ and\ \citenamefont {Vuorinen}}]{Fraga:2013qra}%
  \BibitemOpen
  \bibfield  {author} {\bibinfo {author} {\bibfnamefont {E.~S.}\ \bibnamefont {Fraga}}, \bibinfo {author} {\bibfnamefont {A.}~\bibnamefont {Kurkela}}, \ and\ \bibinfo {author} {\bibfnamefont {A.}~\bibnamefont {Vuorinen}},\ }\href {\doibase 10.1088/2041-8205/781/2/L25} {\bibfield  {journal} {\bibinfo  {journal} {Astrophys. J.}\ }\textbf {\bibinfo {volume} {781}},\ \bibinfo {pages} {L25} (\bibinfo {year} {2014})},\ \Eprint {http://arxiv.org/abs/1311.5154} {arXiv:1311.5154 [nucl-th]} \BibitemShut {NoStop}%
%%CITATION = ARXIV:1311.5154;%%
\bibitem [{Note1()}]{Note1}%
  \BibitemOpen
  \bibinfo {note} {Note that although the pressure is in principle a function of several independent quark chemical potentials, in this work we will consistently parameterize it in terms of the single baryon chemical potential $\mu _B$. The reason for this stems from the fact that with three massless quarks, the physically relevant limits of local charge neutrality and $\beta $-equilibrium are satisfied when $\mu _u=\mu _d=\mu _s=\mu _B/3$. Note, however, that it is trivial to generalize our result to the case of unequal quark chemical potentials, as the new term in the pressure only depends on them via the $m_\infty $ parameter.}\BibitemShut {Stop}%
\bibitem [{\citenamefont {Kurkela}\ \emph {et~al.}(2014)\citenamefont {Kurkela}, \citenamefont {Fraga}, \citenamefont {Schaffner-Bielich},\ and\ \citenamefont {Vuorinen}}]{Kurkela:2014vha}%
  \BibitemOpen
  \bibfield  {author} {\bibinfo {author} {\bibfnamefont {A.}~\bibnamefont {Kurkela}}, \bibinfo {author} {\bibfnamefont {E.~S.}\ \bibnamefont {Fraga}}, \bibinfo {author} {\bibfnamefont {J.}~\bibnamefont {Schaffner-Bielich}}, \ and\ \bibinfo {author} {\bibfnamefont {A.}~\bibnamefont {Vuorinen}},\ }\href {\doibase 10.1088/0004-637X/789/2/127} {\bibfield  {journal} {\bibinfo  {journal} {Astrophys. J.}\ }\textbf {\bibinfo {volume} {789}},\ \bibinfo {pages} {127} (\bibinfo {year} {2014})},\ \Eprint {http://arxiv.org/abs/1402.6618} {arXiv:1402.6618 [astro-ph.HE]} \BibitemShut {NoStop}%
%%CITATION = ARXIV:1402.6618;%%
\bibitem [{\citenamefont {Gorda}(2016)}]{Gorda:2016uag}%
  \BibitemOpen
  \bibfield  {author} {\bibinfo {author} {\bibfnamefont {T.}~\bibnamefont {Gorda}},\ }\href {\doibase 10.3847/0004-637X/832/1/28} {\bibfield  {journal} {\bibinfo  {journal} {Astrophys. J.}\ }\textbf {\bibinfo {volume} {832}},\ \bibinfo {pages} {28} (\bibinfo {year} {2016})},\ \Eprint {http://arxiv.org/abs/1605.08067} {arXiv:1605.08067 [astro-ph.HE]} \BibitemShut {NoStop}%
%%CITATION = ARXIV:1605.08067;%%
\bibitem [{\citenamefont {Annala}\ \emph {et~al.}(2018)\citenamefont {Annala}, \citenamefont {Gorda}, \citenamefont {Kurkela},\ and\ \citenamefont {Vuorinen}}]{Annala:2017llu}%
  \BibitemOpen
  \bibfield  {author} {\bibinfo {author} {\bibfnamefont {E.}~\bibnamefont {Annala}}, \bibinfo {author} {\bibfnamefont {T.}~\bibnamefont {Gorda}}, \bibinfo {author} {\bibfnamefont {A.}~\bibnamefont {Kurkela}}, \ and\ \bibinfo {author} {\bibfnamefont {A.}~\bibnamefont {Vuorinen}},\ }\href {\doibase 10.1103/PhysRevLett.120.172703} {\bibfield  {journal} {\bibinfo  {journal} {Phys. Rev. Lett.}\ }\textbf {\bibinfo {volume} {120}},\ \bibinfo {pages} {172703} (\bibinfo {year} {2018})},\ \Eprint {http://arxiv.org/abs/1711.02644} {arXiv:1711.02644 [astro-ph.HE]} \BibitemShut {NoStop}%
%%CITATION = ARXIV:1711.02644;%%
\bibitem [{\citenamefont {Most}\ \emph {et~al.}(2018)\citenamefont {Most}, \citenamefont {Weih}, \citenamefont {Rezzolla},\ and\ \citenamefont {Schaffner-Bielich}}]{Most:2018hfd}%
  \BibitemOpen
  \bibfield  {author} {\bibinfo {author} {\bibfnamefont {E.~R.}\ \bibnamefont {Most}}, \bibinfo {author} {\bibfnamefont {L.~R.}\ \bibnamefont {Weih}}, \bibinfo {author} {\bibfnamefont {L.}~\bibnamefont {Rezzolla}}, \ and\ \bibinfo {author} {\bibfnamefont {J.}~\bibnamefont {Schaffner-Bielich}},\ }\href {\doibase 10.1103/PhysRevLett.120.261103} {\bibfield  {journal} {\bibinfo  {journal} {Phys. Rev. Lett.}\ }\textbf {\bibinfo {volume} {120}},\ \bibinfo {pages} {261103} (\bibinfo {year} {2018})},\ \Eprint {http://arxiv.org/abs/1803.00549} {arXiv:1803.00549 [gr-qc]} \BibitemShut {NoStop}%
%%CITATION = ARXIV:1803.00549;%%
\bibitem [{\citenamefont {Abbott}\ \emph {et~al.}(2017)\citenamefont {Abbott} \emph {et~al.}}]{TheLIGOScientific:2017qsa}%
  \BibitemOpen
  \bibfield  {author} {\bibinfo {author} {\bibfnamefont {B.}~\bibnamefont {Abbott}} \emph {et~al.} (\bibinfo {collaboration} {Virgo, LIGO Scientific}),\ }\href {\doibase 10.1103/PhysRevLett.119.161101} {\bibfield  {journal} {\bibinfo  {journal} {Phys. Rev. Lett.}\ }\textbf {\bibinfo {volume} {119}},\ \bibinfo {pages} {161101} (\bibinfo {year} {2017})},\ \Eprint {http://arxiv.org/abs/1710.05832} {arXiv:1710.05832 [gr-qc]} \BibitemShut {NoStop}%
%%CITATION = ARXIV:1710.05832;%%
\bibitem [{\citenamefont {Kajantie}\ \emph {et~al.}(2002)\citenamefont {Kajantie}, \citenamefont {Laine},\ and\ \citenamefont {Schroder}}]{Kajantie:2001hv}%
  \BibitemOpen
  \bibfield  {author} {\bibinfo {author} {\bibfnamefont {K.}~\bibnamefont {Kajantie}}, \bibinfo {author} {\bibfnamefont {M.}~\bibnamefont {Laine}}, \ and\ \bibinfo {author} {\bibfnamefont {Y.}~\bibnamefont {Schroder}},\ }\href {\doibase 10.1103/PhysRevD.65.045008} {\bibfield  {journal} {\bibinfo  {journal} {Phys. Rev.}\ }\textbf {\bibinfo {volume} {D65}},\ \bibinfo {pages} {045008} (\bibinfo {year} {2002})},\ \Eprint {http://arxiv.org/abs/hep-ph/0109100} {arXiv:hep-ph/0109100 [hep-ph]} \BibitemShut {NoStop}%
%%CITATION = HEP-PH/0109100;%%
\bibitem [{\citenamefont {Braaten}\ and\ \citenamefont {Pisarski}(1990)}]{Braaten:1989mz}%
  \BibitemOpen
  \bibfield  {author} {\bibinfo {author} {\bibfnamefont {E.}~\bibnamefont {Braaten}}\ and\ \bibinfo {author} {\bibfnamefont {R.~D.}\ \bibnamefont {Pisarski}},\ }\href {\doibase 10.1016/0550-3213(90)90508-B} {\bibfield  {journal} {\bibinfo  {journal} {Nucl. Phys.}\ }\textbf {\bibinfo {volume} {B337}},\ \bibinfo {pages} {569} (\bibinfo {year} {1990})}\BibitemShut {NoStop}%
%%CITATION = NUPHA,B337,569;%%
\bibitem [{\citenamefont {Andersen}\ \emph {et~al.}(1999)\citenamefont {Andersen}, \citenamefont {Braaten},\ and\ \citenamefont {Strickland}}]{Andersen:1999fw}%
  \BibitemOpen
  \bibfield  {author} {\bibinfo {author} {\bibfnamefont {J.~O.}\ \bibnamefont {Andersen}}, \bibinfo {author} {\bibfnamefont {E.}~\bibnamefont {Braaten}}, \ and\ \bibinfo {author} {\bibfnamefont {M.}~\bibnamefont {Strickland}},\ }\href {\doibase 10.1103/PhysRevLett.83.2139} {\bibfield  {journal} {\bibinfo  {journal} {Phys. Rev. Lett.}\ }\textbf {\bibinfo {volume} {83}},\ \bibinfo {pages} {2139} (\bibinfo {year} {1999})},\ \Eprint {http://arxiv.org/abs/hep-ph/9902327} {arXiv:hep-ph/9902327 [hep-ph]} \BibitemShut {NoStop}%
%%CITATION = HEP-PH/9902327;%%
\bibitem [{\citenamefont {Andersen}\ \emph {et~al.}(2000)\citenamefont {Andersen}, \citenamefont {Braaten},\ and\ \citenamefont {Strickland}}]{Andersen:1999sf}%
  \BibitemOpen
  \bibfield  {author} {\bibinfo {author} {\bibfnamefont {J.~O.}\ \bibnamefont {Andersen}}, \bibinfo {author} {\bibfnamefont {E.}~\bibnamefont {Braaten}}, \ and\ \bibinfo {author} {\bibfnamefont {M.}~\bibnamefont {Strickland}},\ }\href {\doibase 10.1103/PhysRevD.61.014017} {\bibfield  {journal} {\bibinfo  {journal} {Phys. Rev.}\ }\textbf {\bibinfo {volume} {D61}},\ \bibinfo {pages} {014017} (\bibinfo {year} {2000})},\ \Eprint {http://arxiv.org/abs/hep-ph/9905337} {arXiv:hep-ph/9905337 [hep-ph]} \BibitemShut {NoStop}%
%%CITATION = HEP-PH/9905337;%%
\bibitem [{\citenamefont {Andersen}\ \emph {et~al.}(2002)\citenamefont {Andersen}, \citenamefont {Braaten}, \citenamefont {Petitgirard},\ and\ \citenamefont {Strickland}}]{Andersen:2002ey}%
  \BibitemOpen
  \bibfield  {author} {\bibinfo {author} {\bibfnamefont {J.~O.}\ \bibnamefont {Andersen}}, \bibinfo {author} {\bibfnamefont {E.}~\bibnamefont {Braaten}}, \bibinfo {author} {\bibfnamefont {E.}~\bibnamefont {Petitgirard}}, \ and\ \bibinfo {author} {\bibfnamefont {M.}~\bibnamefont {Strickland}},\ }\href {\doibase 10.1103/PhysRevD.66.085016} {\bibfield  {journal} {\bibinfo  {journal} {Phys. Rev.}\ }\textbf {\bibinfo {volume} {D66}},\ \bibinfo {pages} {085016} (\bibinfo {year} {2002})},\ \Eprint {http://arxiv.org/abs/hep-ph/0205085} {arXiv:hep-ph/0205085 [hep-ph]} \BibitemShut {NoStop}%
%%CITATION = HEP-PH/0205085;%%
\bibitem [{\citenamefont {Andersen}\ \emph {et~al.}(2004)\citenamefont {Andersen}, \citenamefont {Petitgirard},\ and\ \citenamefont {Strickland}}]{Andersen:2003zk}%
  \BibitemOpen
  \bibfield  {author} {\bibinfo {author} {\bibfnamefont {J.~O.}\ \bibnamefont {Andersen}}, \bibinfo {author} {\bibfnamefont {E.}~\bibnamefont {Petitgirard}}, \ and\ \bibinfo {author} {\bibfnamefont {M.}~\bibnamefont {Strickland}},\ }\href {\doibase 10.1103/PhysRevD.70.045001} {\bibfield  {journal} {\bibinfo  {journal} {Phys. Rev.}\ }\textbf {\bibinfo {volume} {D70}},\ \bibinfo {pages} {045001} (\bibinfo {year} {2004})},\ \Eprint {http://arxiv.org/abs/hep-ph/0302069} {arXiv:hep-ph/0302069 [hep-ph]} \BibitemShut {NoStop}%
%%CITATION = HEP-PH/0302069;%%
\bibitem [{\citenamefont {Haque}\ \emph {et~al.}(2014)\citenamefont {Haque}, \citenamefont {Bandyopadhyay}, \citenamefont {Andersen}, \citenamefont {Mustafa}, \citenamefont {Strickland},\ and\ \citenamefont {Su}}]{Haque:2014rua}%
  \BibitemOpen
  \bibfield  {author} {\bibinfo {author} {\bibfnamefont {N.}~\bibnamefont {Haque}}, \bibinfo {author} {\bibfnamefont {A.}~\bibnamefont {Bandyopadhyay}}, \bibinfo {author} {\bibfnamefont {J.~O.}\ \bibnamefont {Andersen}}, \bibinfo {author} {\bibfnamefont {M.~G.}\ \bibnamefont {Mustafa}}, \bibinfo {author} {\bibfnamefont {M.}~\bibnamefont {Strickland}}, \ and\ \bibinfo {author} {\bibfnamefont {N.}~\bibnamefont {Su}},\ }\href {\doibase 10.1007/JHEP05(2014)027} {\bibfield  {journal} {\bibinfo  {journal} {JHEP}\ }\textbf {\bibinfo {volume} {05}},\ \bibinfo {pages} {027} (\bibinfo {year} {2014})},\ \Eprint {http://arxiv.org/abs/1402.6907} {arXiv:1402.6907 [hep-ph]} \BibitemShut {NoStop}%
%%CITATION = ARXIV:1402.6907;%%
\bibitem [{\citenamefont {Weldon}(1982)}]{Weldon:1982aq}%
  \BibitemOpen
  \bibfield  {author} {\bibinfo {author} {\bibfnamefont {H.~A.}\ \bibnamefont {Weldon}},\ }\href {\doibase 10.1103/PhysRevD.26.1394} {\bibfield  {journal} {\bibinfo  {journal} {Phys. Rev.}\ }\textbf {\bibinfo {volume} {D26}},\ \bibinfo {pages} {1394} (\bibinfo {year} {1982})}\BibitemShut {NoStop}%
%%CITATION = PHRVA,D26,1394;%%
\bibitem [{Note2()}]{Note2}%
  \BibitemOpen
  \bibinfo {note} {Note that this amounts to one further term in the expansion in $m_{D}^{2}$ than what was done in ref.~\cite {Andersen:2002ey}.}\BibitemShut {Stop}%
\bibitem [{Note3()}]{Note3}%
  \BibitemOpen
  \bibinfo {note} {The sign of our result in this equation is incorrect; the correct result is $-1$ times this expression. We thank Jean-Loïc Kneur for bringing this mistake to our attention.}\BibitemShut {Stop}%
\end{thebibliography}%
